\documentclass[prd,preprintnumbers,floatfix,aps,nofootinbib,showpacs]{revtex4} 

\usepackage{latexsym,epsfig,amssymb,amsmath,amstext,nicefrac}

\usepackage{graphicx,floatflt}
\usepackage[usenames]{color}
\usepackage[dvipsnames]{xcolor}

\def\gtrsim{\mathrel{\hbox{\rlap{\hbox{\lower4pt\hbox{$\sim$}}}\hbox{$>$}}}}

\def\lesssim{\mathrel{\hbox{\rlap{\hbox{\lower4pt\hbox{$\sim$}}}\hbox{$<$}}}}

\newcommand{\be}{\begin{equation}}
\newcommand{\ee}{\end{equation}}
\newcommand{\ba}{\begin{eqnarray}}
\newcommand{\ea}{\end{eqnarray}}

\begin{document}

\title{Primordial black holes from single field models of inflation}

\author{Juan Garc\'ia-Bellido}\email{Juan.GarciaBellido@uam.es}
\affiliation{Instituto de F\'isica Te\'orica UAM-CSIC, Universidad Auton\'oma de Madrid,\\ Nicol\'as Cabrera 13, Cantoblanco, 28049  Madrid, Spain\\
CERN, Theoretical Physics Department, 1211 Geneva, Switzerland}

\author{Ester Ruiz Morales} \email{Ester.Ruiz.Morales@upm.es} 
\affiliation{Departamento de F\'isica, Universidad Polit\'ecnica de Madrid,
Ronda de Valencia 3, 28012  Madrid, Spain\\
CERN, Theoretical Physics Department, 1211 Geneva, Switzerland}

\begin{abstract}
Primordial black holes (PBH) have been shown to arise from high peaks in the matter power spectra of multi-field models of  inflation. Here we show, with a simple toy model, that it is also  possible to generate a peak in the curvature power spectrum of  single-field inflation. We assume that the effective dynamics of the  inflaton field presents a near-inflection point which slows down the  field right before the end of inflation and gives rise to a  prominent spike in the fluctuation power spectrum at scales much  smaller than those probed by Cosmic Microwave Background (CMB) and  Large Scale Structure (LSS) observations. This peak will give rise,  upon reentry during the radiation era, to PBH via gravitational  collapse. The mass and abundance of these PBH is such that they  could constitute the totality of the Dark Matter today. We satisfy  all CMB and LSS constraints and predict a very broad range of PBH  masses. Some of these PBH are light enough that they will evaporate  before structure formation, leaving behind a large curvature  fluctuation on small scales. This broad mass distribution of PBH as  Dark Matter will be tested in the future by AdvLIGO and LISA  interferometers. 
\end{abstract}

\pacs{98.80.Cq}

\date{\today}

\preprint{IFT-UAM/CSIC-17-007}

\maketitle

\section{Introduction}\label{section1}
\label{sec:intro}

The nature of Dark Matter (DM) is one of the remaining mysteries of Modern Cosmology. We know it exists since at least 1933, from the first observations of galaxy cluster dynamics by Fritz Zwicky, all the way to full consistency with CMB anisotropies observed by Planck, and LSS formation measured by galaxy surveys. Its nature is still unknown. For a recent review see~\cite{Peebles:2017bzw}. There are more than forty orders of magnitude in mass for the range of possible fundamental constituents, from ultralight axions~\cite{Hui:2016ltb} to supermassive black holes~\cite{Carr:2016drx,Khlopov:2008qy}. Some of the possible Particle Dark Matter (PDM) candidates have recently been strongly constrained as the dominant component of DM (e.g.\ massive neutrinos, WIMPS, heavy axions, etc.), although there is still room for more exotic components.

In this paper we explore the possibility that DM is composed primarily of Primordial Black Holes. The first time a connection between PBH and DM was made was in the paper of Chapline~\cite{Chapline:1975}, where the lightmass PBH of Carr and Hawking~\cite{Carr:1974nx} were supposed to constitute part of the DM. Later on, in 1993, Dolgov and Silk~\cite{Dolgov:1992pu} suggested a model of matter fluctuations which used the QCD transition as a trigger for the formation of PBHs of order a solar mass, which was later developed by Jedamzik~\cite{Jedamzik:1996mr}. These papers relied on the first order nature of the QCD transition to amplify the minute matter fluctuations produced during inflation into high density regions that would gravitationally collapse to form the PBH. We know nowadays that the QCD transition is not first order but a crossover~\cite{Aoki:2006we}, so such a mechanism is ruled out.

An alternative is to produce a large peak in the curvature power spectrum which would collapse to form black holes even without a phase transition. The first time such a peak was used to generate PBH as the main component of DM was proposed by Garc\'ia-Bellido, Linde and Wands in 1996, based on a two-field hybrid model of inflation~\cite{GarciaBellido:1996qt}. Soon afterwards, many papers appeared in the literature making use of peaks in the spectrum~\cite{Yokoyama:1995ex,Nakamura:1997sm,Ivanov:1997ia,Blais:2002nd} to generate a whole range of masses for PBHs as the main constituent of DM. Broad peaks arise from quantum diffusion of the inflaton field that backreact on the metric and induce large curvature perturbations~\cite{Clesse:2015wea,Ivanov:1997ia}. When those fluctuations re-enter the horizon during the radiation era, their gradients induce a gravitational collapse that cannot be overcome even by the radiation pressure of the expanding plasma, producing black holes with a mass of order the horizon mass~\cite{GarciaBellido:1996qt}. Most of these PBH survive until today, and dominate the universe expansion at matter-radiation equality.

In models of inflation with multiple fields, one of the fields acts like the inflaton and the other one either triggers a phase transition or an explosive productions of gauge particles~\cite{Garcia-Bellido:2016dkw} or a fast evolution, all of which backreact on the curvature generating a peak in the spectrum of curvature fluctuations. Depending on when does this phenomenon occur during inflation, i.e.\ how many e-folds before the end of inflation, we may have a narrow or a broad spectrum of masses for the PBHs that form during the radiation era upon re-entry. It is relatively easy in these models to separate the  large scale fluctuations we observe in the CMB and LSS, that correspond to $N=65$--$55$ e-folds, from the small scale fluctuations responsible for PBHs, corresponding to $N\sim40$--$20$ e-folds before the end of inflation.

In this paper we explore single-field models of inflation that may give rise to peaks in the curvature power spectrum responsible for the formation of PBHs. In the absence of a second field that triggers the growth of fluctuations, like in hybrid inflation~\cite{Clesse:2015wea}, we need a period of evolution during inflation where a single field (the inflaton) slows down, creating a stronger backreaction and a quick growth in curvature fluctuations. In terms of dynamics, this effect can only be achieved by producing a plateau in the inflationary potential, which slow-rolls the inflaton even further than usual, before ending inflation. The best way to produce this plateau is with an inflection point in the potential. The problem, in this case, is that the inflaton may stay too long at the inflection point, diluting away the inflationary fluctuations that had successfully imprinted the metric on large scales to explain the CMB anisotropies.

We propose a toy model of single-field inflation, with the minimum number of parameters needed to generate a large peak in the matter power spectrum at small scales, by introducing in the potential a near-inflection point. The main feature of this type of model is that the peak is very broad, and covers many orders of magnitude in mass. In order to account for DM it requires a small fraction $f_{\rm PBH}(M)$ of the total $\Omega_M$ at any given mass interval. This may be the reason why they have not yet been detected by microlensing events or other phenomena that put severe bounds on them. On the other hand, since the peak is broad, it may affect the rate of structure formation since recombination, generating heavy seeds on which supermassive black holes may grow by gas accretion to become the beacons we observe as quasars at high redshift, emitting in X-rays~\cite{Chandra:2017,Luo:2016ojb} and starting early reionization~\cite{Kashlinsky:2016sdv}.

With the advent of Gravitational Wave Astronomy, we have a completely new window into the Early Universe. The fact that PBH are formed in clusters in this scenario~\cite{Clesse:2015wea}, see also~\cite{Chisholm:2005vm,Chisholm:2011kn}, makes them more prone to merging within the age of the universe, and binary inspirals are expected at a higher rate than if they were uniformly distributed in space. For this reason, Clesse and Garc\'ia-Bellido predicted in 2015~\cite{Clesse:2015wea} that LIGO would be able to detect the final merging of massive BH binaries (BHB), as it actually occurred a few months later~\cite{Abbott:2016blz}. The large masses of the progenitors (36 and 29 solar masses) prompted the attention of the community, which immediately suggested a connection between Dark Matter and PBH~\cite{Bird:2016dcv,Clesse:2016vqa,Sasaki:2016jop}, and initiated a search for a variety of formation mechanisms to be explored, including both primordial~\cite{Kawasaki:2016pql,Carr:2016drx,Nakama:2016kfq,Garcia-Bellido:2016dkw,Blinnikov:2016bxu,Erfani:2015rqv,Deng:2016vzb} and via stellar evolution~\cite{Raccanelli:2016cud,Belczynski:2016jno}. A few months later we learned that there had been three BHB merger events in LIGO Run O1~\cite{Abbott:2016nmj,TheLIGOScientific:2016pea,Abbott:2016bqf}. Furthermore, last week the LIGO collaboration announced two more events. The surprisingly high rate of merger events suggests that black holes are more abundant and more clustered that expected, in agreement with the PBH scenario.

Moreover, not only do we have individual BHB merger events, but soon we will be able to detect the Stochastic GW Background from such inspirals in LIGO~\cite{TheLIGOScientific:2016wyq}, happening all the way from recombination until today, if they arise from PBHs~\cite{Mandic:2016lcn,Clesse:2016ajp}. Moreover, in the near future, the space interferometer LISA should have this as a new irreducible background~\cite{Bartolo:2016ami}, and will be able to measure the main parameters of the PBH mass distribution.

The paper is structured as follows.  In section~\ref{section2} we introduce the single-field toy model of inflation and obtain exact expressions for the slow-roll parameters and the number of e-folds of inflation. We also study the validity of the slow-roll approximation for different values of the model 
parameters. In section~\ref{section3} we compute the curvature power spectrum for this simple model at all scales, comparing the amplitude at large scales (CMB and LSS) with that at small scales (responsible for PBH). In section~\ref{section4} we study the production of PBH and their evaporation, and we conclude with an overall discussion of the scenario in section~\ref{section5}.

\section{Single-field toy model of inflation}\label{section2}
\label{sec:toymodel}

Although several multi-field models of inflation have been proposed that give rise to peaks in the matter power spectrum, e.g.\ by having a mild waterfall regime at the transition between slow-roll and the symmetry breaking stage of hybrid inflation~\cite{Clesse:2015wea}, no compelling single-field model of inflation exists that can give a significant spike in the matter power spectrum of fluctuations.

In this paper we show that a single-field scalar potential with a near-inflection point can also give rise to a peak in the matter power spectrum. This simple and quite generic feature of the potential makes the inflaton enter into an ultra-slow-rolling stage during a short range of e-folds that generates the spike in the spectrum. One has to be careful however that this new feature in the potential does not affect the already constrained spectrum on CMB and LSS scales. In multi-field models like hybrid inflation this can easily be achieved, since one can decouple the two stages (CMB and PBH) by the intermediate waterfall stage, which arises due to the coupling to the symmetry breaking field. As we will see, in single field models of inflation, those two stages are intimately related, imposing much tighter constraints on the model parameters.

We propose a single-field toy model constructed as a ratio of polynomials, in order to take advantage of the asymptotic flatness of the potential for large values of the field, which seems to be in agreement with CMB observations. This type of potentials arise naturally in models of Higgs Inflation~\cite{Bezrukov:2007ep,GarciaBellido:2008ab}. However, in this case, rather than starting below (but close to) the inflection point, as usually done in MSSM inflation~\cite{Allahverdi:2006iq,Allahverdi:2006we}, in Accidental Inflation~\cite{Linde:2007jn} and in more general Inflection-point Inflation~\cite{Hotchkiss:2011am}, we will assume that the field starts at larger values,\footnote{It is not clear, however, that our toy model with two plateaus can be embedded in a natural way within MSSM or Accidental Inflation.} well above the inflection point, and then slow-rolls down towards the minimum of the potential, crossing the near-inflection point a few (typically 30 to 40) e-folds before the end of inflation.

\begin{figure}[t]
\centering
\includegraphics[width=0.7\textwidth]{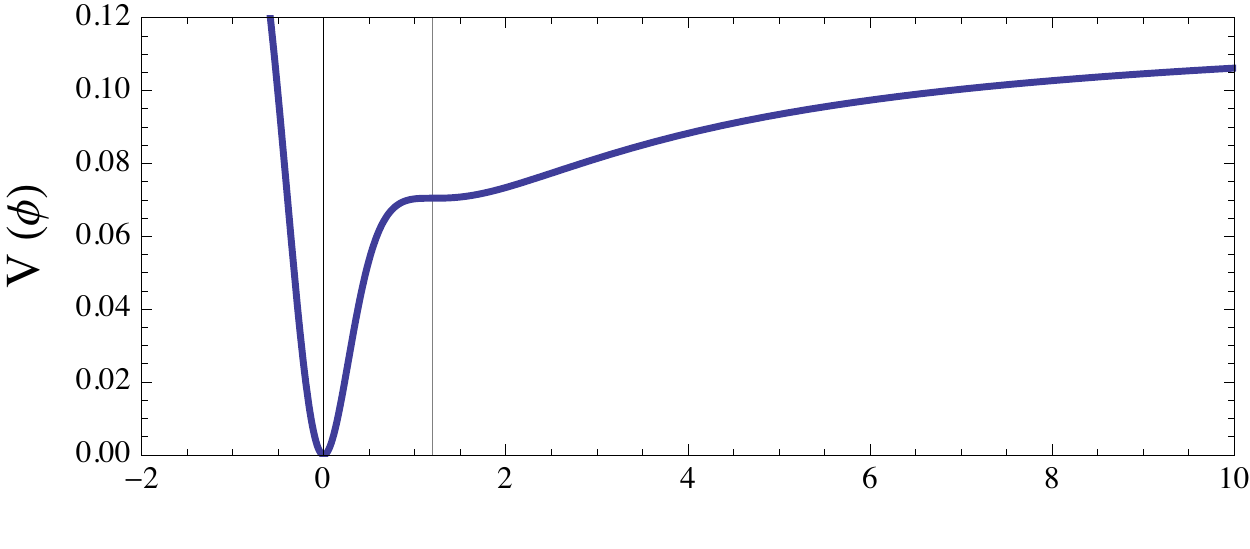}\vspace*{-7mm}
\includegraphics[width=0.7\textwidth]{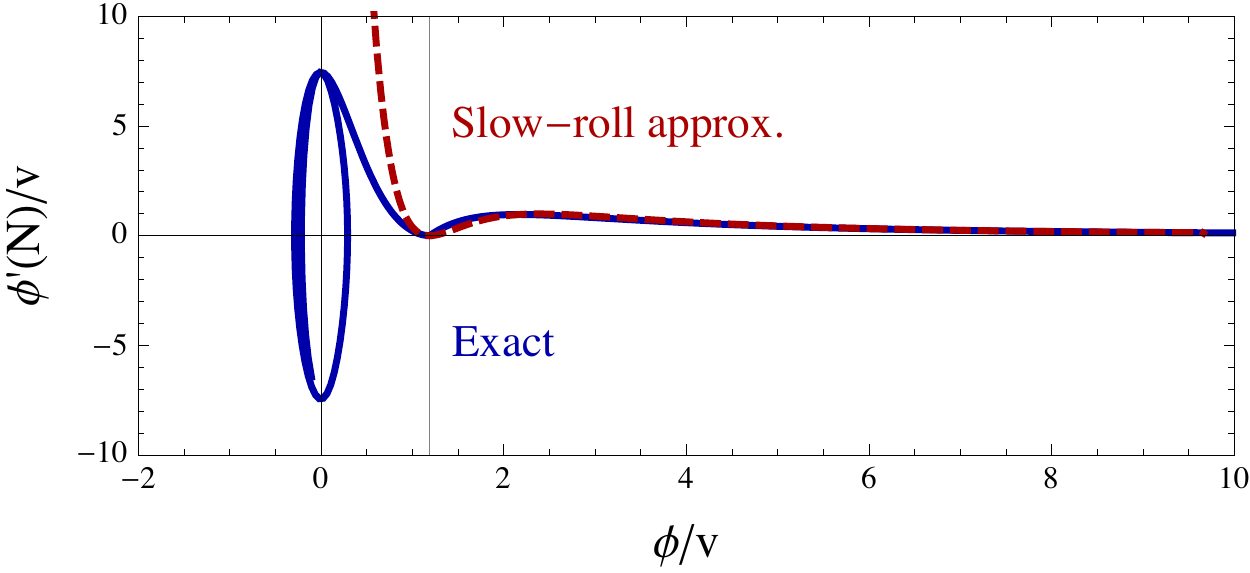}
\caption{Top figure: Single field potential $V(\phi)$ (arbitrary units) with an inflection point (vertical line) at $\phi=1.191\,v$, and an asymptotically flat plateau. The model parameters are $a=1,\, b=b_c(1)-\beta,\, \beta=1\times10^{-4}$ and $\kappa^2v^2 = 0.108$. Bottom figure: The exact evolution in phase space of the inflaton field (blue) for the same potential, compared with the slow-roll approximation (dashed red).}
\label{fig:potential}
\end{figure}

The inflationary potential we propose as toy-model is given by
\be
V(\phi) = \left(\frac{1}{2} m^2\,\phi^2 - \frac{1}{3}\alpha\,v\,\phi^3 + \frac{1}{4}\lambda\,\phi^4\right)
\Big(1+\xi\,\phi^2\Big)^{-2}\,,
\label{eq:pot}
\ee
which can be recast in the form
\be
V(x) = \frac{\lambda\,v^4}{12}\ \frac{x^2(6 - 4\,a\,x + 3\,x^2)}{(1+b\,x^2)^2}\,,
\label{eq:pot2}
\ee
under a redefinition of parameters, $x=\phi/v$, $m^2=\lambda\,v^2$, $a=\alpha/\lambda$ and $b=\xi\,v^2$.

This potential has extrema, $V'(\phi)=0$, for certain values of the parameters $(a,\, b)$ which can be obtained by solving the third order equation
\be
1 - a\,x+(1-b)\,x^2+\frac{ab}{3}\,x^3=0\,,
\label{eq:thirdorder}
\ee

The three solutions of eq.~(\ref{eq:thirdorder}) are given by $x=x_1<0$ and $x_{2,\,3} = x_0 \pm i\,y_0$, with
\begin{align}
x_1  & =   \frac{b-1}{a\,b} - \frac{1}{a\,b}\left(\Theta(a,\,b) + \frac{(b-1)^2+a^2\,b}{\Theta(a,\,b)}\right),\\
x_0  & =   \frac{b-1}{a\,b} + \frac{1}{2\,a\,b}\left(\Theta(a,\,b) +
\frac{(b-1)^2+a^2\,b}{\Theta(a,\,b)}\right),\\
y_0  & =   \frac{\sqrt3}{2\,a\,b}\left(\Theta(a,\,b) - \frac{(b-1)^2+a^2\,b}{\Theta(a,\,b)}\right),
\end{align}
and
\be
\Theta(a,\,b) = \left(\sqrt3\,a\,b\,\sqrt{\Delta(a)^3 -
\left(b -1+\frac{1}{3}a^2\right)^3} +
(1-b)^3+\frac{3}{2}\,a^2\,b\right)^{1/3}\,.
\ee

From these expressions we see that the potential will have an inflection point $V''(x)=0$ at a real value of $x$, whenever the parameter $b$ acquires a critical value (as a function of $a$)~\footnote{Note that there is a maximum value of $b_c$ (for $a=1/\sqrt2$) and also a maximum value of $a$ ($a=2$) for which the critical $b$ parameter is positive.}
\begin{align}
b_c(a)   &=  1 - \frac{1}{3}\,a^2 + \Delta(a) \,, \\
\Delta(a)  &=  \frac{a^2}{3}\,\left(\frac{9}{2a^2}-1\right)^{2/3} \,.
\end{align}
\looseness=-1 
For these critical values, $y_0=0$,  the solutions of the cubic equation simplify to
\begin{alignat}{2}
x_1 &= \frac{b_c - 1 - 2\sqrt{(b_c-1)^2\!+\!a^2\,b_c}}{a\,b_c} &&< 0\,, 
\label{eq:x1}\\
x_0 &= \frac{b_c - 1 + \sqrt{\,(b_c-1)^2+a^2\,b_c}}{a\,b_c} &&> 0\,. 
\label{eq:x2}
\end{alignat}
\looseness=-1 and the potential has an inflection point at  $x=x_{2,3}=x _0$ given by eq.~(\ref{eq:x2}).
For values of the parameter $b<b_c$, $y_0$ is real, i.e.\ there is only one real solution $x_1$, and the potential has no inflection point. In the case that $b>b_c$, $y_0$ is pure imaginary, the three solutions are real, and the inflection point breaks up into a maximum and a minimum at $x_2,\, x_3$.

Although the potential has four independent parameters $(\lambda,\, v, \, a, \, b)$,  the existence of  an inflection point only depends on $(a,\,b)$. We have plotted the potential in figure~\ref{fig:potential}a, for a representative choice of the parameters, $a=1$ and $b=b_c(1) = 1.435$, that we will use along the paper, and for which the inflection point is at small values of the field, $\phi=1.191\,v$. 

\subsection{Slow-roll parameters and number of e-folds}\label{section2.1}

\begin{figure}[t]
\centering\hspace*{-3mm}
\includegraphics[width=0.51\textwidth]{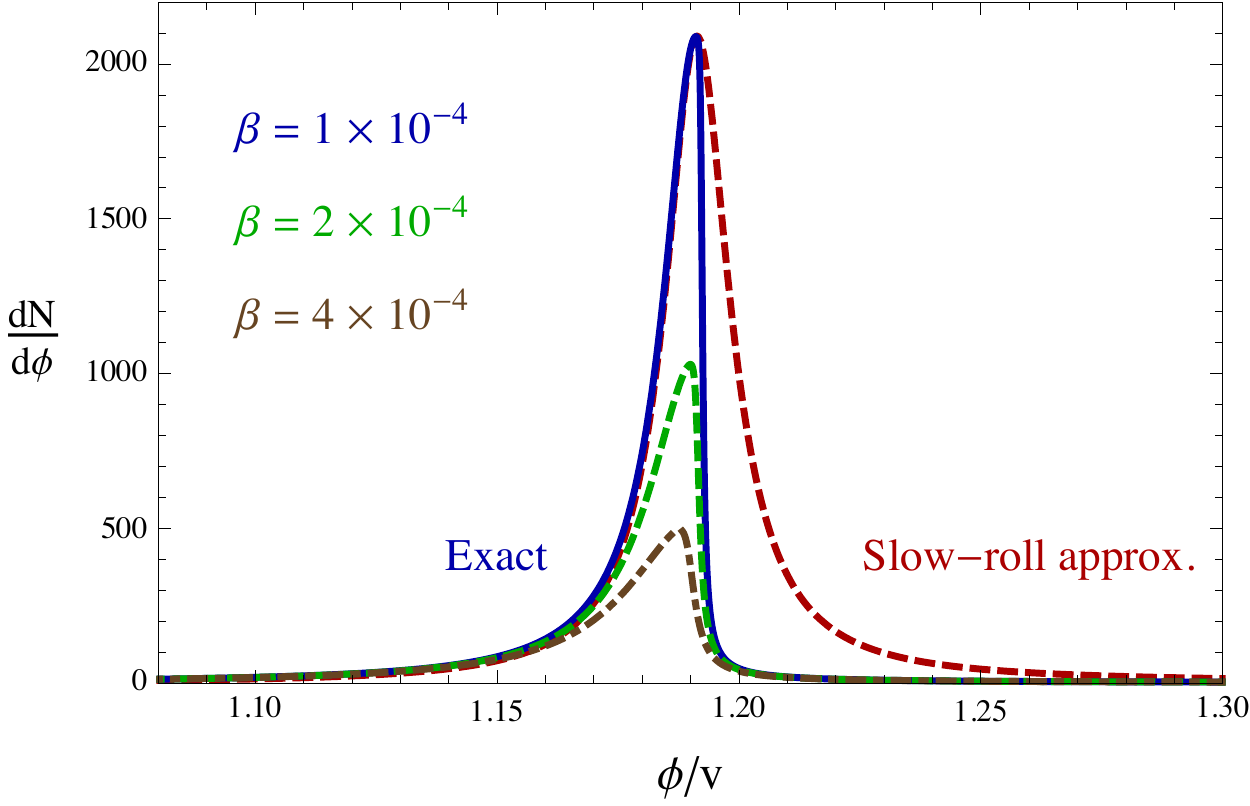}
\includegraphics[width=0.475\textwidth]{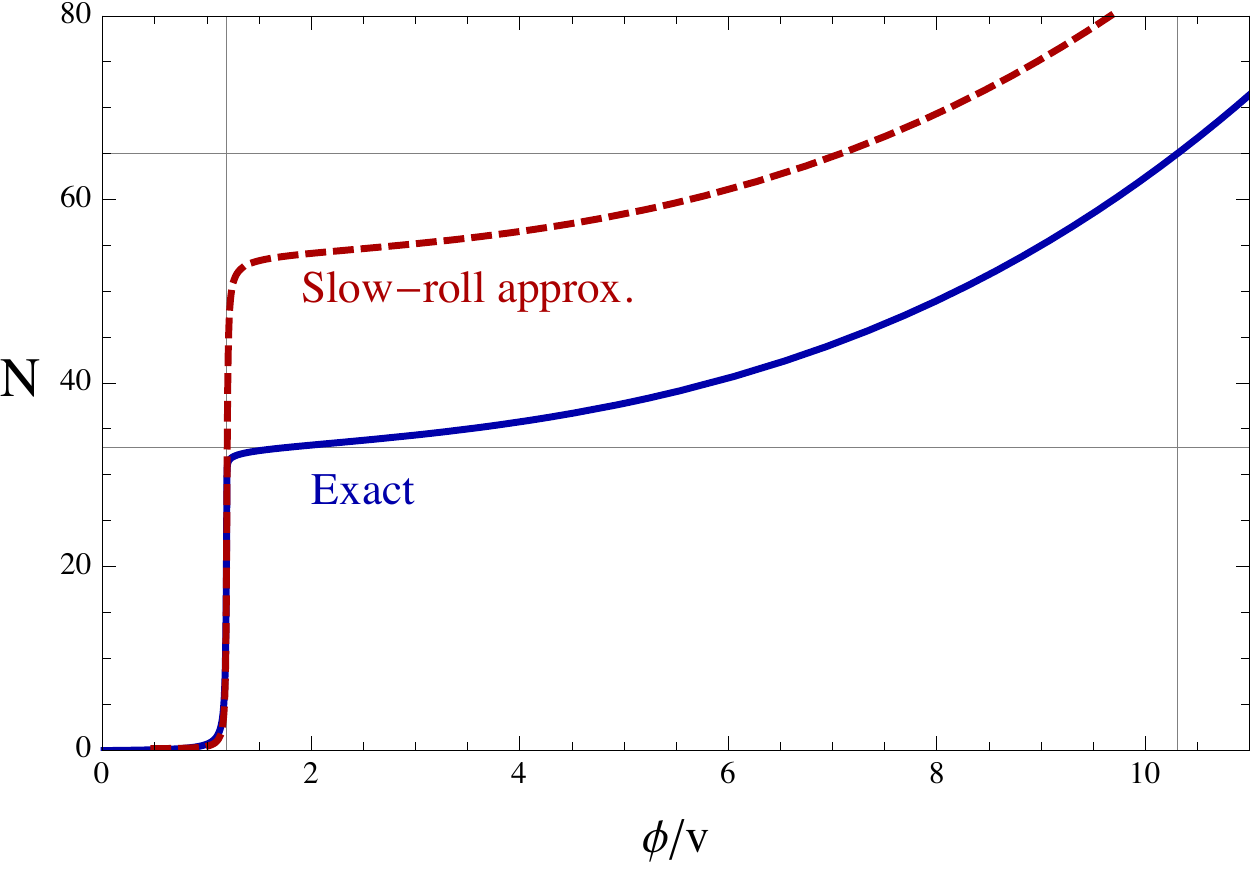}
\caption{ Left panel: The integrand of the number of e-folds as a function of $\phi/v$. The dashed-red curve corresponds to the SRA with $\beta=1\times10^{-4}$. The exact calculation (blue, green and brown curves) is also shown for three different values of the resonance parameter $\beta=(1,\,2,\,4)\times10^{-4}$. Right panel: The number of e-folds of inflation $N$ as a function of $\phi/v$, in the SRA (dashed-red) and the exact calculation (blue). In both figures, the model parameters are $a=1,\, b=b_c(1)-\beta$ and $\kappa^2v^2 = 0.108$.
\label{fig:dNdx}}
\end{figure}

In this type of potential  the slow-roll parameter $\epsilon$ can be calculated
in the slow-roll approximation (SRA) as
\be\label{eq:eps}
\epsilon_{\rm SR} = \frac{1}{2\kappa^2}\left(\frac{V'(\phi)}{V(\phi)}\right)^2
= \frac{8}{\kappa^2 v^2}\ \frac{(3-3\,a\,x+3\,(1-b)\,x^2+a\,b\,x^3)^2}{x^2\, (6-4\,a\,x+3\,x^2)^2\, (1+b\,x^2)^2}\,,
\ee
and the number of e-folds is given by
\begin{align}\nonumber
N_{\rm SR} = \int \frac{\kappa d\phi}{\sqrt{2\epsilon_{\rm SR}}}
 & = \frac{\kappa^2 v^2}{4} \int \frac{x\,(6-4\,a\,x+3\,x^2)\,(1+b\,x^2)}{(3-3\,a\,x+3\,(1-b)\,x^2+a\,b\,x^3)}\,dx\,, \\[2mm] \label{eq:Nefolds}
 &=  \frac{\kappa^2 v^2}{4\,a\,b} \int \frac{x\,(6-4\,a\,x+3\,x^2)\,(1+b\,x^2)}{(x-x_1)\,((x-x_0)^2+y_0^2)}\,dx\,.
\end{align}
\looseness=-1 Clearly, for model parameters $(a,\,b)$ in their critical values, the inflaton passes through the inflection point ($x\to x_0$, $y_0 \to 0$), and the integrand $dN/dx$ in eq.~(\ref{eq:Nefolds}) diverges in the SRA. However, for values sufficiently close to the critical ones $(a,\, b=b_c(a)-\beta)$ with the resonance parameter $0<\beta\ll1$, the potential has a \emph{near-inflection point}. In that case, for each value of $(a,\, b_c(a))$, and by appropriately choosing the parameters $\beta$ and $\kappa v$, we can have a large number of e-folds spent at $x=x_0$ to produce a significant peak in the matter power spectrum.
In figure~\ref{fig:dNdx}, we have plotted as a dashed-red line the integrand of eq.~(\ref{eq:Nefolds}) in the SRA to show the Breit-Wigner shape of the resonance at $x=x_0$. 

In recent papers~\cite{Kannike:2017bxn,Germani:2017bcs,Motohashi:2017kbs,Dimopoulos:2017ged}, the use of the slow-roll approximation has been discussed in potentials with an inflection point. In order to clarify this issue in our model, we have 
numerically integrated the evolution of the inflaton field $\phi$ following the \emph{exact} equations:
\begin{equation}\label{eq:FiedEq}
\ddot\phi + 3 H \dot\phi + V'(\phi)   =   0\,, \hspace{1cm}
H^2 =  \frac{\kappa^2}{3} \left( \frac{1}{2} \dot\phi^2 +V(\phi) \right)\,,
\hspace{1cm} \dot H =  - \frac{\kappa^2}{2} \dot\phi^2,
\end{equation} 
and compared it with the slow-roll approximation. In figure~\ref{fig:potential}(b), we show the evolution of the inflaton field in phase space. It is clear that the slow-roll trajectory follows quite closely the exact result of the field equations~(\ref{eq:FiedEq}) until the end of inflation.\footnote{Note also that initial conditions are irrelevant since the inflaton is always in the SR attractor at large values of $\phi$.}
The solution shown corresponds to parameter values chosen along this paper $a=1$, $b=b_c(1)-\beta$, $\beta=1\times10^{-4}$ and $\kappa^2v^2=0.108$. It is possible to obtain similar trajectories (with large number of e-folds spent at the near-inflection point before ending inflation) for any values of $a\in(0,2)$, and $b=b_c(a) - \beta$, as long as the parameter $\kappa^2v^2$ is appropriately chosen.

\begin{figure}[t]
\centering
\vspace{-5pt}
\includegraphics[width=0.47\textwidth]{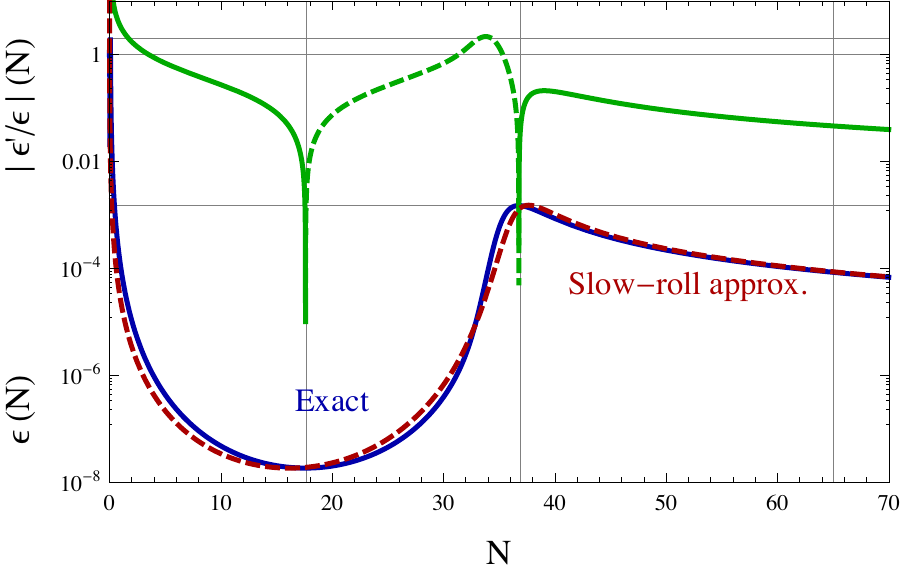}\hspace{5mm}
\includegraphics[width=0.47\textwidth]{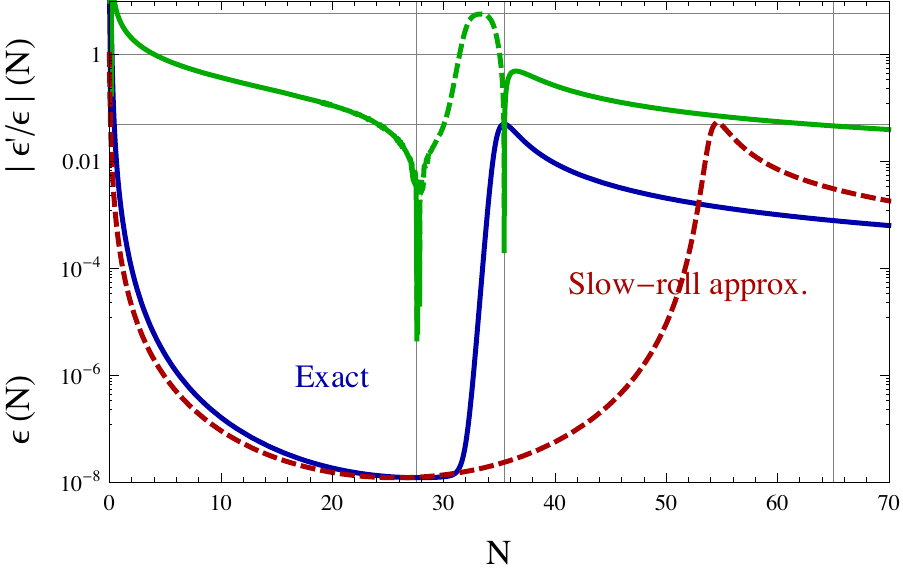}
\caption{The \emph{exact} parameter $\epsilon'(N)/\epsilon(N)$ (in green,  dashed section corresponds to negative values), together with the $\epsilon(N)$ parameter in the exact calculation (in blue) and SRA (in red), along the evolution during inflation. Fig.~(a) corresponds to $a=0.2$, $b=b_c(0.2)-\beta$, $\beta=10^{-4}$ and $\kappa^2v^2=0.00333$, while Fig.~(b) is for our representative parameter choice, $a=1$, $b=b_c(1)-\beta$, $\beta=10^{-4}$ and $\kappa^2v^2=0.108$. Both choices give $\Delta N=35$ and $n_s=0.954$ on CMB scales.
}
\label{fig:SRA}
\end{figure}

Once we have solved the exact evolution of $\phi$, with Eqs.~(\ref{eq:FiedEq}), the slow-roll parameter $\epsilon$ and the total number of e-folds of inflation $N$ can be calculated \emph{exactly}:
\begin{equation} \label{eq:slpar}
\epsilon(\phi)  =   - \frac{\dot H}{H^2}\,, \hspace{1cm}
N(\phi) = \int_{\phi_{\rm end}}^\phi  \frac{\kappa d\phi}{\sqrt{2\epsilon(\phi)}}\,.
\end{equation} 

In order to investigate the validity of the slow-roll approximation for different choices of the model parameters, we have shown in figure~\ref{fig:SRA} the exact evolution of the $\epsilon$ parameter and
its logarithmic derivative, as a function of the number of e-folds, compared with $\epsilon_{\rm SR}$ in the SRA (\ref{eq:eps}). We have chosen two representative cases ($a=1$ and $a=0.2$).
In both cases, $\epsilon$ is always well below $0.05$, throughout the whole inflationary evolution, and the parameter $\epsilon'/\epsilon$ only touches the value $-3$ at an instant (for the $a=1$ case) before the inflection point, and does not become permanently trapped in an ultra-slow-roll regime.

Scanning the allowed range of values of the parameter $a$, we have confirmed that the SRA becomes less and less accurate as we increase $a$, although this does not prevent the slow-roll approximation from capturing the main features of our model: growth of $N$ at the quasi-inflection point, and enhanced power spectra at small scales, as we will see in the next section. It is a general feature of the model that, for those cases in which the SRA deviates from the exact evolution of the $\epsilon$ parameter, see figure~\ref{fig:SRA}(b), the SRA always overestimates the number of e-folds spent at the inflection point, as shown in figure~\ref{fig:dNdx}.

Furthermore, note that, contrary to what was claimed in~\cite{Germani:2017bcs}, the number of e-folds spent at the inflection point, $\Delta N$ strongly depends on the value of the resonant parameter $\beta$, both in the exact and the SRA. The closer we are to the critical value of $b$, the larger the number of e-folds spent at the critical point. This can be seen in figure~\ref{fig:dNdx}, where the area under the curve gives a number of e-folds spent close to the resonance, $\Delta N = 35, \, 22, \, 16$, in the exact calculation, for three choices of $\beta$.

\section{Power spectrum from single field inflation}\label{section3}
\label{sec:power}

In this section we compute the curvature power spectrum and make sure that a peak at small scales will not be in conflict with the  good properties of the spectrum at CMB and LSS scales. The exact expression of the power spectrum is given by
\be\label{eq:Pkexact}
P_{\cal R}(k) = \frac{\kappa^2H^2(\phi)}{8\pi^2\epsilon(\phi)}\,,
\ee
while in the SRA the power spectrum can be evaluated in terms of the potential~(\ref{eq:pot2}) and the slow-roll parameter~(\ref{eq:eps}) as
\be\label{eq:Pk}
P_{\cal R}^{\rm SR}(k) = \frac{\kappa^4 \,V(x)}{24\pi^2\,\epsilon_{\rm SR}(x)} =
\frac{\lambda\,\kappa^6v^6}{96\times24\pi^2}\ \frac{(6 - 4\,a\,x + 3\,x^2)^3\,x^4}
{(3-3\,a\,x+3\,(1-b)\,x^2+a\,b\,x^3)^2}\,.
\ee

Note that, since the spectrum is essentially inversely proportional to the parameter $\epsilon$, for large values of $a$ in which the SR estimate for $\epsilon$ can significantly deviate from the exact result (see figure~\ref{fig:SRA}), the power spectrum calculated with eq.~(\ref{eq:Pk}) can give a very different prediction from the exact result (\ref{eq:Pkexact}), as can be explicitly seen in Fig.~\ref{fig:PRk}.

It is convenient to study the behavior of the power spectrum at different asymptotic regimes, at small and large scales, in order to check its compatibility with existing CMB and LSS bounds on large scales, and with the amount of PHB produced at small scales.

\subsection{$P_{\cal R}(k)$ at large scales: CMB and LSS}\label{section3.1}

\begin{figure}[t]
\centering
\includegraphics[width=0.75\textwidth]{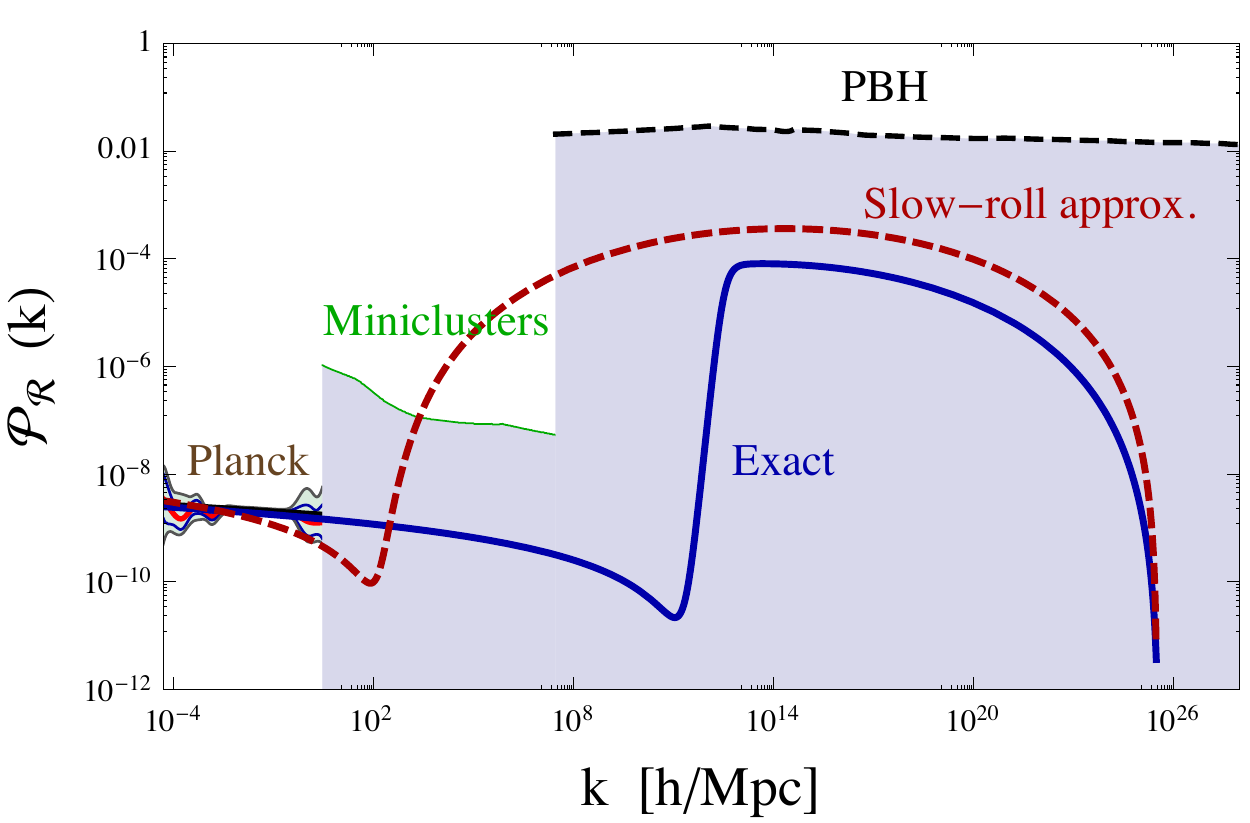}
\caption{
The exact matter power spectrum (blue line), for model parameters:
$a=1$,
$b=b_c(1)-\beta$,
$\kappa^2 v^2 = 0.108$ and
$\beta = 1\times10^{-4}$. We have also
plotted the SRA (red-dashed line), and the range
of values allowed by Planck (2015), by compact minihalos (green line) and by PBH (black dashed line), at 95\% c.l. (figure adapted from ref.~\cite{Bringmann:2011ut}).\label{fig:PRk}}
\end{figure}

We compute the relevant CMB parameters, i.e. the scalar spectral index and its running, as well as the tensor-to-scalar ratio~\cite{Garcia-Bellido:2014gna}, using the exact expressions for $\epsilon(N)$ from eq.~(\ref{eq:slpar}) as
\begin{align}\label{ns}
n_s  = \frac{d\ln P_{\cal R}}{d\ln k} &=  1 - 2\,\epsilon(N) + \frac{\epsilon'(N)}{\epsilon(N)} \,,\\[2mm]  \label{dns}
\frac{d n_s}{d\ln k}  &=   2\,\epsilon'(N) - \left(\frac{\epsilon'(N)}{\epsilon(N)}\right)' \,, \\[3mm] \label{eq:r}
r  &=  16\,\epsilon(N) \,.
\end{align}

We will only consider model parameters that are compatible with the Planck measurements at $k=0.05$ Mpc$^{-1}$~\cite{Ade:2015lrj,Ade:2015xua}, 
\begin{alignat}{2}\label{eq:Planck}
\ln(10^{10}\,A_s^2)  &=  3.094 \pm 0.068& \hspace{1.5cm} &(95 \%\ {\rm c.l.})\,, \\[2mm]
n_s  &=  0.9569 \pm 0.0154 &\hspace{1.1cm} &(95 \%\ {\rm c.l.})\,, \\
\frac{d\,n_s}{d\ln k}   &=  0.011 \pm 0.028 &\hspace{1.5cm} &(95 \%\ {\rm c.l.})\,, \\[1mm]
r  &<  0.09 & \hspace{3.1cm} &(95 \%\ {\rm c.l.})\,.
\end{alignat}

There is a whole range of model parameters that satisfy these constraints. The amplitude of the power spectrum at Planck scales is given by $A_s^2 = P_{\cal R}(x_{62})$, where $x_{62}$ is the inflaton value (in units of $v$) 62 e-folds before the end of inflation, which in our model corresponds to $k=0.05$ Mpc$^{-1}$. One of the nice features of the potential of our model (\ref{eq:pot}) is its flatness at large values of the inflaton field, since this allows for a reasonable value of the spectral index and its running. For our choice of values of the model parameters, $a=1$, $b=b_c(1)-\beta$, $\kappa^2 v^2 = 0.108$ and \mbox{$\beta = 1\times10^{-4}$}, we find
\begin{align}
n_s &= 0.951\,, \\[2mm]
\frac{d\,n_s}{d\ln k} &= -0.00175\,, \\[3mm]
r &=0.013\,, 
\end{align}
which are in agreement with CMB constraints. Finally, the amplitude~(\ref{eq:Planck}) fixes the quartic self-coupling,  $\lambda= 2.97 \times 10^{-7}$. With these parameters, the exact power spectrum satisfies all the observational constraints in the whole range of scales, from both CMB and LSS, as can be seen in figure~\ref{fig:PRk}. Note, however, that the slow-roll approximation would have predicted the wrong spectral index and a ${\cal P}_{\cal R}(k)$ already ruled out by experiment, see red-dashed curve in figure~\ref{fig:PRk}. Therefore, we agree with ref.~\cite{Germani:2017bcs} that in realistic models of inflation with a quasi-inflection point, a careful calculation of the \emph{exact} evolution has to be performed, in order to get accurate predictions for the power spectrum.

\begin{figure}[t]
\centering
\includegraphics[width=0.75\textwidth]{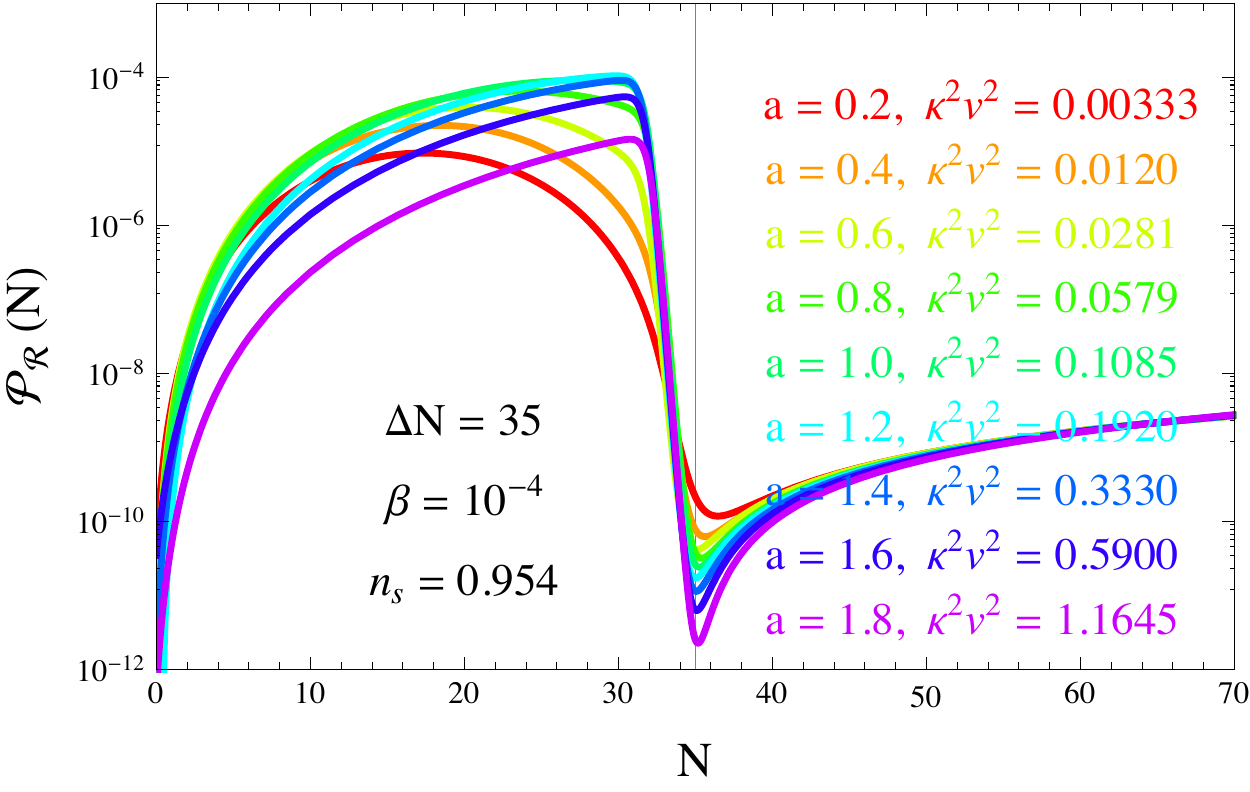}
\caption{The curvature power spectra obtained from the exact inflaton evolution, for different values of the model parameters ($a,\,\kappa^2v^2$) for which there is a significant peak at small scales, while leaving CMB scales essentially unaffected, with $\Delta N=35$ and $n_s=0.954$. In all cases $\beta = 10^{-4}$.}
\label{fig:PkN}
\end{figure}

Finally, it is important to stress here, following the discussion of the previous section,  that the agreement with the CMB at large scales is a generic feature of this model, that can be obtained independently of the value of $a$, as long as we chose the appropriate scale $v$. In figure~\ref{fig:PkN} we show the exact calculation of the power spectra of our model for the whole range of parameters ($a,\, \kappa^2v^2$), which are in agreement with CMB observations, and at the same time present a peak at small scales, as we discuss in the next subsection.

\subsection{$P_{\cal R}(k)$ at small scales: PBH}\label{section3.2}

Since we are interested in producing a population of primordial black holes, once we have chosen the model parameters to satisfy CMB constraints on large scales, our model has to be able to generate also high peaks in the power spectrum at small scales. This effect can be achieved dynamically via a secondary plateau in the inflationary potential, which slow-rolls the inflaton even further than usual, increasing the local number of e-folds spent near the quasi-inflection point (see figure \ref{fig:dNdx}), before ending inflation. 

In our model, the exact power spectrum~(\ref{eq:Pkexact}) presents a peak at scales corresponding to the near-inflection point $(x=x_0)$. 
In figure~\ref{fig:PkN} we have covered the whole parameter range ($a,\,\kappa^2v^2$) and shown that such a peak is generic, although its shape and amplitude depends on the choice of model parameters. Note that for small values of $a$, for which the SRA is valid, the peak has a arc shape, while for larger values of $a$ it acquires a half-dome shape, because the $\epsilon$ parameter changes more abruptly in these cases. This feature will be useful for PBH production since it shifts the peak mass in the BH distribution to larger values.

Once we have fixed all the model parameters, we have to check that these gives rise to reasonable physical values of the scale of inflation and the reheating temperature. The value of $\kappa^2 v^2 = 0.108$ implies that we have inflation significantly below the Planck scale, and that we are thus safe from quantum gravity corrections. Moreover, since the mass of the inflaton at the minimum of the potential is relatively high, $m = v\sqrt\lambda =4.36 \times10^{14}$\,GeV, and the energy density at the end of inflation is $\rho_{\rm end} = \frac{3}{2}V(\phi_{\rm end})=5.06 \times10^{63}$\,GeV$^4$, we expect a relatively high reheating temperature,
$T_{\rm rh} =3.46 \times10^{15}$\,GeV, and thus a large total number of e-folds, $N=65$.

\section{Production of primordial black holes}\label{section4}
\label{sec:PBH}

\looseness=-1 Here we study how the peak in the matter power spectrum generates a peak in the mass distribution of black holes and how to compute their abundance as a fraction of the total $\Omega_M$.

Assuming that the probability distribution of density perturbations are Gaussian, one can evaluate the fraction $\beta^{\rm form}$ of the universe collapsing into primordial black holes of mass $M$ at the time of formation $t_M$ as~\cite{Clesse:2015wea}
\be
\beta^{\rm form}(M)  \equiv   \left. \frac{\rho_{\rm PBH} (M) }{\rho_{\rm tot}} \right|_{t=t_M}
  =   \int_{\zeta_c}^{\infty} \frac{{\rm d} \zeta}{\sqrt{2 \pi} \sigma}\, e^{- \frac{\zeta^2}{2 \sigma^2}}  =  \frac{1}{2} \, {\rm erfc} \left( \frac{\zeta_c}{\sqrt 2 \sigma}  \right) \,.
\ee
In the limit where $\sigma \ll \zeta_c$, one gets
\be \label{eq:betaform}
\beta^{\rm form}(M) = \frac{ \sigma}{ \sqrt{2\pi}\, \zeta_c} \, e^{- \frac{\zeta_c^2}{2 \sigma^2 }}.
\ee
The variance of the curvature perturbations $\sigma$  is related to the power spectrum through $\langle \zeta^2 \rangle = \sigma^2 = \mathcal P_\zeta (k_M) $, where $k_M$ is the wavenumber of the mode re-entering inside the Hubble radius at time $t_M$.  If $\sigma$ is above a certain threshold, $\zeta_c \sim 0.05 - 1$, then the probability of collapse to form a PBH can be large~\cite{Clesse:2015wea}. The actual value of $\zeta_c$ depends very much on the equation of state of the fluid at reentry~\cite{Harada:2013epa}
\be
\zeta_c = \left.\frac{1}{3}\ln\frac{3(\chi_a-\sin\chi_a\cos\chi_a)}{2\sin^3\chi_a}\right|_{\chi_a = \pi\sqrt w/(1+3w)}\,,
\ee
which for the radiation epoch, $w=1/3$, gives $\zeta_c = 0.086$.\footnote{Although larger estimates have been found in previous studies~\cite{Green:2004wb}, the precise value is not crucial for our analysis since we can always adjust the $\beta$ parameter correspondingly, to ensure a sufficiently large number density of PBH at formation.} The mass of the PBH at formation is then given by
\be
M_{\rm PBH} = \gamma \frac{4\pi M_P^2}{H_N}\,e^{2N}\,,
\ee
where $\gamma \sim 0.4$ is an efficiency factor characterizing the gravitational collapse upon reentry of the metric fluctuations responsible for the PBH formation~\cite{Green:2004wb,Musco:2004ak}, and $H_N$ is the rate of expansion during inflation at the time of horizon exit, while $N$ is the number of e-folds when the fluctuation exits during inflation. For fluctuations produced during the near-inflection point, this rate is given (e.g.\ for the model parameters
$a=1$,
$b=b_c(1)-\beta$, $\beta = 1\times10^{-4}$ and
$x_0=1.191$) by
\be
H_N =1.8 \times 10^{-6} M_P\,.
\ee
Putting all together, one finds the mass of PBHs, in solar mass units, as a function of $N$,
\be\label{eq:MPBH}
M_{\rm PBH} = M_\odot\,e^{2(N-36.28)}\,.
\ee

%\subsection{PBH evaporation in broad mass distributions}\label{section4.1}

When PBH form, their whole distribution cover a wide range of masses, some of which will evaporate immediately, much before nucleosynthesis, while others will remain until matter radiation equality and recombination, and yet another group of more massive PBH will remain today to form the Dark Matter.

\begin{figure}
\centering
\includegraphics[width=0.75\textwidth]{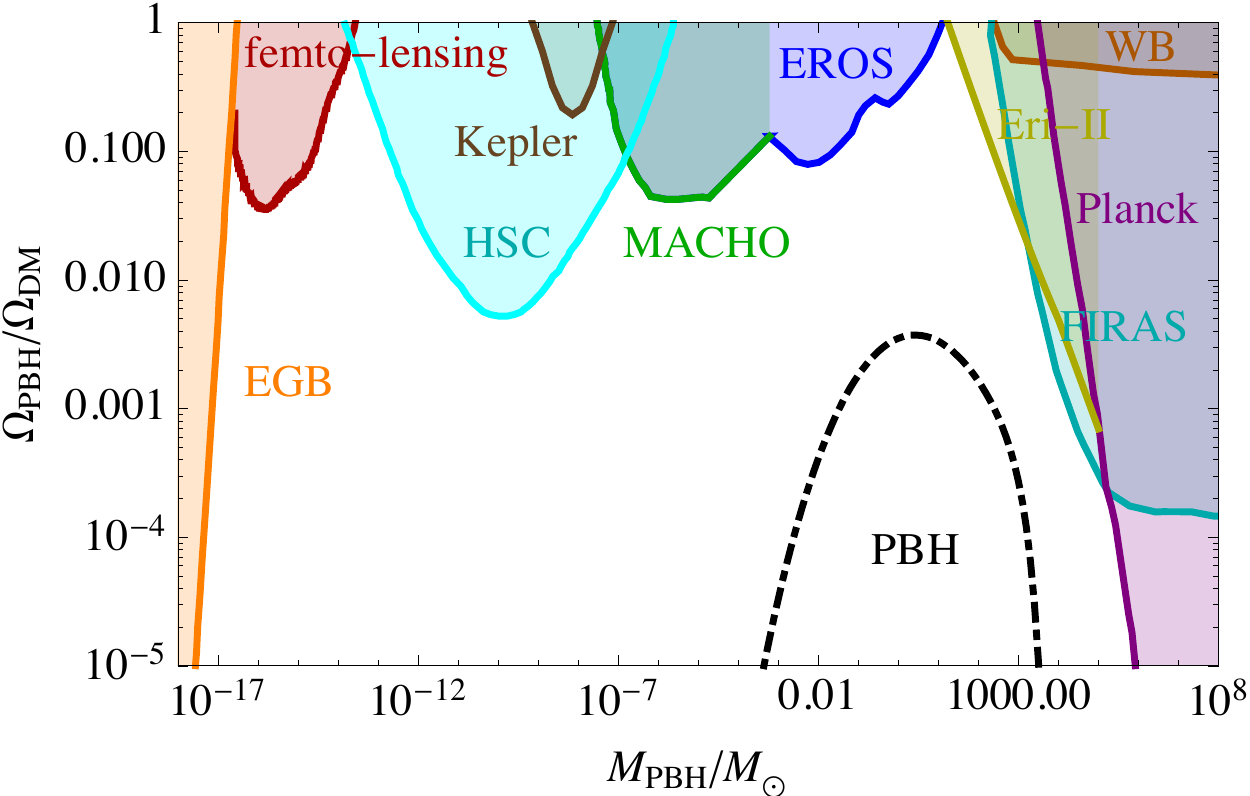}
\caption{The present constraints on PBH from Extragalactic Gamma Background, femto-mili-microlensing, Wide Binaries and the CMB.
The primordial
black holes produced in this type of near-inflection-point inflationary models could comprise all of the dark matter and still pass all the constraints, see the text. Figure adapted from~\cite{Clesse:2015wea} and~\cite{Carr:2016drx}.\label{fig:constPBH}}
\end{figure}

The evaporation via Hawking radiation is a runaway process that lasts a fraction of a second for small BHs, but can take much more than the age of the universe for massive ones,
\be\label{eq:tev}
\tau_{\rm ev} = 5120\,\pi\,\sqrt{\frac{\hbar G}{c^5}}\left(\frac{M_{\rm PBH}}{M_P}\right)^3 =
2.1\times 10^{67} \ {\rm yr} \,\left(\frac{M_{\rm PBH}}{M_\odot}\right)^3 =
13.8 \ {\rm Gyr} \ e^{6(N-17.21)}\,.
\ee
Their relative abundance today will depend on their initial mass~(\ref{eq:MPBH}) and their relative abundance~(\ref{eq:betaform}). Those that do not evaporate during the radiation era will come to dominate the density of the universe at matter-radiation equality. Their abundance will grow relative to that of radiation since formation to equality by a factor~\cite{Clesse:2015wea}
\be\label{eq:betaeq}
\beta^{\rm eq}(M) = \frac{a_{\rm eq}}{a(t_M)}\,\beta^{\rm form}(M)\,,
\ee
and their contribution to the matter content of the universe at equality is
\be\label{eq:OmegaM}
\Omega_{\rm PBH}^{\rm eq} = \int_{M^*}^{M_{\rm eq}} \frac{dM}{M}\,\beta^{\rm eq}(M) =
\int_{12.51}^{38} 2\,d N\,e^{56.8-N}\,\beta^{\rm form}(N)\,,
\ee
where $56.8=\ln(T_{\rm rh}/T_{\rm eq})$ is the number of e-folds from the end of inflation to equality, and $M^* \simeq 3\times10^{12}\ {\rm g}$ is the lower mass limit of those PBH that have survived until equality without evaporating. The final number, $\Omega_{\rm PBH}^{\rm eq} =0.42$, corresponds to the case where these PBH constitute all of the DM today (the rest of matter at equality, $\Omega_B^{\rm eq}=0.08$, corresponds to baryons). For the parameters of the model, $a=1$, $b=b_c(1)-\beta$, $\kappa^2 v^2 = 0.108$ and $\beta = 1\times10^{-4}$, we have found that $\zeta_c=0.0753$ gives rise to $\Omega_{\rm PBH}^{\rm eq} =0.42$ at equality. Note that a growth in mass of PBH since formation, due to accretion before equality, would increase $\Omega_{\rm PBH}^{\rm eq}$ for a fixed $\zeta_c$; for example, one could have obtained $\Omega_{\rm PBH}^{\rm eq} =0.42$ for larger $\zeta_c$ (e.g. $\sim0.4$) by taking into account accretion and the growth of perturbations during the radiation era~\cite{Garcia-Bellido:2017aan}. For simplicity, in this paper we have only considered its evolution after equality, assuming a factor $2\times10^7$ increase in mass due to PBH merging.\footnote{In Refs~\cite{Chisholm:2005vm,Chisholm:2011kn} it is argued that the concentration (density contrast) of PBH could be enhanced by huge factors (up to $10^{22}$). Part of it is simply the reduction of the size of the cluster through dynamical friction, and part is due to internal collisions and merging of PBH, which increases their mass and shifts the whole mass distribution to larger masses. A detailed analysis of this non-linear phenomenon requires N-body simulations of clusters of PBH during the matter era, in the presence of large curvature fluctuations, which are now been explored by one the authors (work in progress).} The effect of gas accretion in the matter era is not taken into account, but it is expected to significantly raise the tail of the distribution for larger masses.

We show in figure~\ref{fig:constPBH} the present constraints on PBH coming from the evaporation of micro black holes inducing an Extragalactic Gamma Background, from femto-mili-microlensing (Kepler, MACHO, EROS, OGLE), from Wide Binaries and the CMB (FIRAS, WMAP), see~\cite{Carr:2016drx} for the latest compilation. Note that the constraints shown in Fig.~\ref{fig:constPBH} correspond to a monocromatic mass distribution. More recently, Ref~\cite{Carr:2017jsz} studied the effect of wide mass distributions and found, for a distribution similar to ours (see their Fig.~1b for a lognormal with $\sigma=2$), a slight broadening and depletion of the constraints. Since our normalised PBH distribution is well below all the experimental bounds, by at least an order of magnitude, the broadening of the constraints does not pose a significant restriction on the model.

\section{Discussion and conclusions}\label{section5}
\label{sec:conclusions}

We have shown that it is possible to construct single-field models of inflation that satisfy all the constraints on large scales (from the CMB to compact minihalos LSS) and at the same time provide a mechanism for generating the observed Dark Matter in the form of PBH. The main advantage of single-field models is the economy of means. What multi-field models like hybrid inflation or axion inflation generate with new couplings, can be done here with a plateau feature in the single-scalar-field potential. We have succeeded in producing a peak in the primordial curvature power spectrum with just two parameters that govern the width and height of the inflationary plateau. While most models of inflation are only probed at CMB and LSS scales, i.e.\ approximately 60 to 50 e-folds before the end of inflation, the range of values of the field relevant for PBH production and the CMB covers the whole range, from 60 to the end of inflation. What is surprising is that generating a peak in the spectrum did not require any special fine-tuning of parameters for the amplitude of CBM fluctuations; actually, for our choice of parameters, the inflaton self-coupling $\lambda$ turns out to be of order $10^{-7}$, and the inflaton mass is of order the GUT scale.

The fact that quantum fluctuations of the inflaton field can backreact on space-time and form classical inhomogeneities, giving rise to CMB anisotropies and big structures like galaxies and clusters, is a fascinating property of Quantum Field Theory in curved space, and one of the great successes of inflation. What is even more surprising is that a local feature in the inflaton dynamics can give rise to large amplitude fluctuations in the spatial curvature, which collapse to form black holes upon reentry during the radiation era, and that these PBH may constitute today the bulk of the matter in the universe. In this new scenario, Dark Matter is no longer a particle produced after inflation, whose interactions must be deduced from high energy particle physics experiments, but an object formed by the gravitational collapse of relativistic particles in the early universe when subject to high curvature gradients, themselves produced by quantum fluctuations of a field and stretched to cosmological scales by inflation.

This opens a new window into the Early Universe. Cosmological observations of PBH as the main component of dark matter in the present universe, mainly through the characterization of their mass distribution, may give us information about the inflaton dynamics just a few tens of e-folds before the end of inflation. Moreover, this new paradigm has many observational consequences~\cite{Garcia-Bellido:2017fdg}. For instance, this scenario could explain the missing satellite problem of N-body simulations, and makes specific predictions for the existence of massive black holes at the center of \emph{all} large scale structures, from dwarf spheroidals to massive galaxies~\cite{Li:2016utv}. The hierarchical merger tree scenario of structure formation then predicts that dark matter halos are composed of an intermediate mass black hole at their center, plus a smooth component of lighter PBH orbiting the halos. These building blocks will then merge to form larger structures like galaxies and clusters of galaxies.

What characterizes this scenario of inflation with broad peaks in $P_{\cal R}(k)$~\cite{Clesse:2015wea,Garcia-Bellido:2016dkw} is the small scale structure they predict. Rather than waiting for $5\sigma$ peaks in a Gaussian low-amplitude spectrum to gravitationally collapse to form the first stars after recombination, here the large-amplitude peaks in the primordial spectrum are enough to produce the PBH that will act as seeds on which gas will accrete and form galaxies. This scenario is thus responsible for reionization at high redshifts and early structure formation, which may explain why we observe fully formed structures like galaxies and QSO so far back in time, very soon after photon decoupling, and why there are strong spatial correlations between the cosmic near IR background and the soft-X-ray background fluctuations~\cite{Kashlinsky:2016sdv}.

Furthermore, in this single-field model the economy of parameters imply that high curvature peaks cover a broad range of scales, and thus give rise to PBH with a large range of masses. The evaporation of these light PBH, between their time of formation and matter-radiation equality, is a new feature of this model. The rest of massive PBH will constitute the DM\@. Note that the large metric (curvature) perturbations remain there on small scales even if the PBH that formed by them upon reentry have evaporated today. Those large fluctuations will grow since equality and recombination due to the gravitational collapse of gas and PBH as DM around them, and will generate structures on extremely small scales, thus acting as early seeds for galaxies.

Fortunately, with the advent of Gravitational Wave Astronomy we have a completely new way of investigating both the late and the early universe. The gravitational wave background from the time of formation of PBH will leave signatures on PTA and LISA scales, while the inspiralling of PBHs to form more massive ones can be seen as chirps in both LISA and AdvLIGO-VIRGO. The stochastic background from these inspirals will cover the whole range from nHz to kHz, and in the not so far future, LISA and the Einstein Telescope will be able to detect such a background and hopefully determine the shape of the mass distribution.

We have entered a new era, with a new scenario of structure formation based on compact BH as the dominant component of Dark Matter, and new detectors in the form of laser interferometers, both on the ground and in space. We will be able to use these new tools to explore the formation of PBH in the early universe and connect these with the dynamics of inflation at energy scales close to the scale of quantum gravity, where present particle physics accelerators cannot reach.

\acknowledgments

We thank Andrei Linde, Sebastien Clesse and Yashar Akrami for useful comments and discussions.  We thank the hospitality of the Lorentz Institute in Leiden during the THCOS@LN meeting and specially Ana Achucarro for her generosity. We also thank the Theory Department at CERN for their hospitality during our Sabbatical year at CERN. This work is supported by the Research Project of the Spanish MINECO, FPA2015-68048-C3-3-P (MINECO-FEDER), and the Centro de Excelencia Severo Ochoa Program SEV-2012-0597. JGB acknowledges support from the Salvador de Madariaga Program Ref. PRX17/00056. ERM acknowledges support from a Research Mobility Grant from the Programa Propio UPM.


\begin{thebibliography}{99}

%\cite{Peebles:2017bzw}
\bibitem{Peebles:2017bzw} 
  P.~J.~E.~Peebles,
  ``How the Nonbaryonic Dark Matter Theory Grew,''
  arXiv:1701.05837 [astro-ph.CO].
  %%CITATION = ARXIV:1701.05837;%%


%\cite{Hui:2016ltb}
\bibitem{Hui:2016ltb} 
  L.~Hui, J.~P.~Ostriker, S.~Tremaine and E.~Witten,
  %``Ultralight scalars as cosmological dark matter,''
  Phys.\ Rev.\ D {\bf 95}, 043541 (2017),
  %doi:10.1103/PhysRevD.95.043541
  [arXiv:1610.08297 [astro-ph.CO]].
  %%CITATION = doi:10.1103/PhysRevD.95.043541;%%


%\cite{Carr:2016drx}
\bibitem{Carr:2016drx} 
  B.~Carr, F.~Kuhnel and M.~Sandstad,
  %``Primordial Black Holes as Dark Matter,''
  Phys.\ Rev.\ D {\bf 94}, 083504 (2016)
  %doi:10.1103/PhysRevD.94.083504
  [arXiv:1607.06077 [astro-ph.CO]].
  %%CITATION = doi:10.1103/PhysRevD.94.083504;%%
  %37 citations counted in INSPIRE as of 10 Feb 2017


%\cite{Khlopov:2008qy}
\bibitem{Khlopov:2008qy} 
  M.~Y.~Khlopov,
  %``Primordial Black Holes,''
  Res.\ Astron.\ Astrophys.\  {\bf 10}, 495 (2010),
  %doi:10.1088/1674-4527/10/6/001
  [arXiv:0801.0116 [astro-ph]].
  %%CITATION = doi:10.1088/1674-4527/10/6/001;%%
  %97 citations counted in INSPIRE as of 19 Feb 2017


\bibitem{Chapline:1975}
 G.~F.~Chapline,
 ``Cosmological effects of primordial black holes,"
 Nature {\bf 253}, 251 (1975) 
 %doi:10.1038/253251a0

%\cite{Carr:1974nx}
\bibitem{Carr:1974nx} 
  B.~J.~Carr and S.~W.~Hawking,
  %``Black holes in the early Universe,''
  Mon.\ Not.\ Roy.\ Astron.\ Soc.\  {\bf 168}, 399 (1974).
  %%CITATION = MNRAA,168,399;%%
  %411 citations counted in INSPIRE as of 12 Feb 2017


%\cite{Dolgov:1992pu}
\bibitem{Dolgov:1992pu} 
  A.~Dolgov and J.~Silk,
  %``Baryon isocurvature fluctuations at small scales and baryonic dark matter,''
  Phys.\ Rev.\ D {\bf 47}, 4244 (1993).
  %doi:10.1103/PhysRevD.47.4244
  %%CITATION = doi:10.1103/PhysRevD.47.4244;%%
  %102 citations counted in INSPIRE as of 10 Feb 2017

%\cite{Jedamzik:1996mr}
\bibitem{Jedamzik:1996mr} 
  K.~Jedamzik,
  %``Primordial black hole formation during the QCD epoch,''
  Phys.\ Rev.\ D {\bf 55}, 5871 (1997)
  %doi:10.1103/PhysRevD.55.5871
  [astro-ph/9605152].
  %%CITATION = doi:10.1103/PhysRevD.55.5871;%%
  %114 citations counted in INSPIRE as of 10 Feb 2017

 %\cite{Aoki:2006we}
\bibitem{Aoki:2006we} 
  Y.~Aoki, G.~Endrodi, Z.~Fodor, S.~D.~Katz and K.~K.~Szabo,
  %``The Order of the quantum chromodynamics transition predicted by the standard model of particle physics,''
  Nature {\bf 443}, 675 (2006),
  %doi:10.1038/nature05120
  [hep-lat/0611014].
  %%CITATION = doi:10.1038/nature05120;%%
  %915 citations counted in INSPIRE as of 12 Feb 2017

  
%\cite{GarciaBellido:1996qt}
\bibitem{GarciaBellido:1996qt} 
  J.~Garc\'ia-Bellido, A.~D.~Linde and D.~Wands,
  %``Density perturbations and black hole formation in hybrid inflation,''
  Phys.\ Rev.\ D {\bf 54}, 6040 (1996)
  %doi:10.1103/PhysRevD.54.6040
  [astro-ph/9605094].
  %%CITATION = doi:10.1103/PhysRevD.54.6040;%%
  %240 citations counted in INSPIRE as of 10 Feb 2017


%\cite{Yokoyama:1995ex}
\bibitem{Yokoyama:1995ex} 
  J.~Yokoyama,
  %``Formation of MACHO primordial black holes in inflationary cosmology,''
  Astron.\ Astrophys.\  {\bf 318}, 673 (1997)
  [astro-ph/9509027].
  %%CITATION = ASTRO-PH/9509027;%%
  %82 citations counted in INSPIRE as of 10 Feb 2017

%\cite{Nakamura:1997sm}
\bibitem{Nakamura:1997sm} 
  T.~Nakamura, M.~Sasaki, T.~Tanaka and K.~S.~Thorne,
  %``Gravitational waves from coalescing black hole MACHO binaries,''
  Astrophys.\ J.\  {\bf 487}, L139 (1997)
  %doi:10.1086/310886
  [astro-ph/9708060].
  %%CITATION = doi:10.1086/310886;%%
  %91 citations counted in INSPIRE as of 10 Feb 2017

%\cite{Ivanov:1997ia}
\bibitem{Ivanov:1997ia} 
  P.~Ivanov,
  %``Nonlinear metric perturbations and production of primordial black holes,''
  Phys.\ Rev.\ D {\bf 57}, 7145 (1998)
  doi:10.1103/PhysRevD.57.7145
  [astro-ph/9708224].
  %%CITATION = doi:10.1103/PhysRevD.57.7145;%%
  %56 citations counted in INSPIRE as of 12 Sep 2017

%\cite{Blais:2002nd}
\bibitem{Blais:2002nd} 
  D.~Blais, C.~Kiefer and D.~Polarski,
  %``Can primordial black holes be a significant part of dark matter?,''
  Phys.\ Lett.\ B {\bf 535}, 11 (2002)
  doi:10.1016/S0370-2693(02)01803-8
  [astro-ph/0203520].
  %%CITATION = doi:10.1016/S0370-2693(02)01803-8;%%
  %52 citations counted in INSPIRE as of 12 Sep 2017


 %\cite{Clesse:2015wea}
\bibitem{Clesse:2015wea} 
  S.~Clesse and J.~Garc\'ia-Bellido,
  %``Massive Primordial Black Holes from Hybrid Inflation as Dark Matter and the seeds of Galaxies,''
  Phys.\ Rev.\ D {\bf 92}, 023524 (2015)
  %doi:10.1103/PhysRevD.92.023524
  [arXiv:1501.07565].
  %%CITATION = doi:10.1103/PhysRevD.92.023524;%%
  %16 citations counted in INSPIRE as of 29 Sep 2016


%\cite{Garcia-Bellido:2016dkw}
\bibitem{Garcia-Bellido:2016dkw} 
  J.~Garc\'ia-Bellido, M.~Peloso and C.~Unal,
  %``Gravitational waves at interferometer scales and primordial black holes in axion inflation,''
  JCAP {\bf 1612}, 031 (2016),
  %doi:10.1088/1475-7516/2016/12/031
  [arXiv:1610.03763].
  %%CITATION = doi:10.1088/1475-7516/2016/12/031;%%
  %7 citations counted in INSPIRE as of 10 Feb 2017


%\cite{Chandra:2017}
\bibitem{Chandra:2017}
 F.~Vito {\it et al.} [Chandra Collaboration],
 %The deepest X-ray view of high-redshift galaxies: constraints on low-rate black-hole accretion,"
 Mon.\ Not.\ Roy.\ Astron.\ Soc.\  {\bf 463}, 348 (2016),
 %doi:10.1093/mnras/stw1998
 [arXiv:1608.02614 [astro-ph.GA]],

%\cite{Luo:2016ojb}
\bibitem{Luo:2016ojb} 
  B.~Luo {\it et al.} [Chandra Collaboration],
  %``The Chandra Deep Field-South Survey: 7 Ms Source Catalogs,''
  Astrophys.\ J.\ Suppl.\  {\bf 228}, 2 (2017),
  %doi:10.3847/1538-4365/228/1/2
  [arXiv:1611.03501 [astro-ph.GA]].
  %%CITATION = doi:10.3847/1538-4365/228/1/2;%%
  %2 citations counted in INSPIRE as of 10 Feb 2017


%\cite{Kashlinsky:2016sdv}
\bibitem{Kashlinsky:2016sdv} 
  A.~Kashlinsky,
  %``LIGO gravitational wave detection, primordial black holes and the near-IR cosmic infrared background anisotropies,''
  Astrophys.\ J.\  {\bf 823}, L25 (2016),
  %doi:10.3847/2041-8205/823/2/L25
  [arXiv:1605.04023 [astro-ph.CO]].
  %%CITATION = doi:10.3847/2041-8205/823/2/L25;%%
  %11 citations counted in INSPIRE as of 10 Feb 2017

% PBH CLUSTERING AND MERGING

%\cite{Chisholm:2005vm}
\bibitem{Chisholm:2005vm} 
  J.~R.~Chisholm,
  %``Clustering of primordial black holes: basic results,''
  Phys.\ Rev.\ D {\bf 73}, 083504 (2006),
  %doi:10.1103/PhysRevD.73.083504
  [astro-ph/0509141].
  %%CITATION = doi:10.1103/PhysRevD.73.083504;%%
  %25 citations counted in INSPIRE as of 19 Feb 2017

%\cite{Chisholm:2011kn}
\bibitem{Chisholm:2011kn} 
  J.~R.~Chisholm,
  %``Clustering of Primordial Black Holes. II. Evolution of Bound Systems,''
  Phys.\ Rev.\ D {\bf 84}, 124031 (2011),
  %doi:10.1103/PhysRevD.84.124031
  [arXiv:1110.4402 [astro-ph.CO]].
  %%CITATION = doi:10.1103/PhysRevD.84.124031;%%
  %3 citations counted in INSPIRE as of 19 Feb 2017


%  FIRST DETECTION OF GW BY LIGO

%\cite{Abbott:2016blz}
\bibitem{Abbott:2016blz} 
  B.~P.~Abbott {\it et al.} [LIGO Scientific and Virgo Collaborations],
  %``Observation of Gravitational Waves from a Binary Black Hole Merger,''
  Phys.\ Rev.\ Lett.\  {\bf 116}, 061102 (2016),
  %doi:10.1103/PhysRevLett.116.061102
  [arXiv:1602.03837 [gr-qc]].
  %%CITATION = doi:10.1103/PhysRevLett.116.061102;%%
  %706 citations counted in INSPIRE as of 28 Sep 2016


%  EARLY CLAIMS OF CONNECTION  GW - PBH
  
 %\cite{Bird:2016dcv}
\bibitem{Bird:2016dcv} 
  S.~Bird, I.~Cholis, J.~B.~Mu\~noz, Y.~Ali-Ha\"imoud, M.~Kamionkowski, E.~D.~Kovetz, A.~Raccanelli and A.~G.~Riess,
  %``Did LIGO detect dark matter?,''
  Phys.\ Rev.\ Lett.\  {\bf 116}, 201301 (2016),
  %doi:10.1103/PhysRevLett.116.201301
  [arXiv:1603.00464 [astro-ph.CO]].
  %%CITATION = doi:10.1103/PhysRevLett.116.201301;%%
  %38 citations counted in INSPIRE as of 04 Oct 2016

 
 %\cite{Clesse:2016vqa}
\bibitem{Clesse:2016vqa} 
  S.~Clesse and J.~Garc\'ia-Bellido,
  %``The clustering of massive Primordial Black Holes as Dark Matter: measuring their mass distribution with Advanced LIGO,''
  Phys.\ Dark Univ.\  {\bf 10}, 002 (2016),
%  doi:10.1016/j.dark.2016.10.002
  [arXiv:1603.05234].
  %%CITATION = doi:10.1016/j.dark.2016.10.002;%%
  %37 citations counted in INSPIRE as of 17 Jan 2017

 
%\cite{Sasaki:2016jop}
\bibitem{Sasaki:2016jop} 
  M.~Sasaki, T.~Suyama, T.~Tanaka and S.~Yokoyama,
  %``Primordial Black Hole Scenario for the Gravitational-Wave Event GW150914,''
  Phys.\ Rev.\ Lett.\  {\bf 117}, 061101 (2016),
  %doi:10.1103/PhysRevLett.117.061101
  [arXiv:1603.08338 [astro-ph.CO]].
  %%CITATION = doi:10.1103/PhysRevLett.117.061101;%%
  %26 citations counted in INSPIRE as of 06 Oct 2016


% PRODUCTION MECHANISMS


%\cite{Kawasaki:2016pql}
\bibitem{Kawasaki:2016pql} 
  M.~Kawasaki, A.~Kusenko, Y.~Tada and T.~T.~Yanagida,
  %``Primordial black holes as dark matter in supergravity inflation models,''
  Phys.\ Rev.\ D {\bf 94}, 083523 (2016),
  %doi:10.1103/PhysRevD.94.083523
  [arXiv:1606.07631 [astro-ph.CO]].
  %%CITATION = doi:10.1103/PhysRevD.94.083523;%%
  %9 citations counted in INSPIRE as of 10 Feb 2017

%\cite{Nakama:2016kfq}
\bibitem{Nakama:2016kfq} 
  T.~Nakama, T.~Suyama and J.~Yokoyama,
  %``Supermassive black holes formed by direct collapse of inflationary perturbations,''
  Phys.\ Rev.\ D {\bf 94}, 103522 (2016),
  %doi:10.1103/PhysRevD.94.103522
  [arXiv:1609.02245 [gr-qc]].
  %%CITATION = doi:10.1103/PhysRevD.94.103522;%%
  %2 citations counted in INSPIRE as of 11 Feb 2017

%\cite{Blinnikov:2016bxu}
\bibitem{Blinnikov:2016bxu} 
  S.~Blinnikov, A.~Dolgov, N.~K.~Porayko and K.~Postnov,
  %``Solving puzzles of GW150914 by primordial black holes,''
  JCAP {\bf 1611}, 036 (2016),
  %doi:10.1088/1475-7516/2016/11/036
  [arXiv:1611.00541 [astro-ph.HE]].
  %%CITATION = doi:10.1088/1475-7516/2016/11/036;%%
  %1 citations counted in INSPIRE as of 11 Feb 2017

%\cite{Erfani:2015rqv}
\bibitem{Erfani:2015rqv} 
  E.~Erfani,
  %``Primordial Black Holes Formation from Particle Production during Inflation,''
  JCAP {\bf 1604}, 020 (2016),
  %doi:10.1088/1475-7516/2016/04/020
  [arXiv:1511.08470 [astro-ph.CO]].
  %%CITATION = doi:10.1088/1475-7516/2016/04/020;%%
  %2 citations counted in INSPIRE as of 19 May 2016

%\cite{Deng:2016vzb}
\bibitem{Deng:2016vzb} 
  H.~Deng, J.~Garriga and A.~Vilenkin,
  ``Primordial black hole and wormhole formation by domain walls,''
  arXiv:1612.03753 [gr-qc].
  %%CITATION = ARXIV:1612.03753;%%
  %1 citations counted in INSPIRE as of 12 Feb 2017

%\cite{Raccanelli:2016cud}
\bibitem{Raccanelli:2016cud} 
  A.~Raccanelli, E.~D.~Kovetz, S.~Bird, I.~Cholis and J.~B.~Munoz,
  %``Determining the progenitors of merging black-hole binaries,''
  Phys.\ Rev.\ D {\bf 94}, 023516 (2016),
  %doi:10.1103/PhysRevD.94.023516
  [arXiv:1605.01405 [astro-ph.CO]].
  %%CITATION = doi:10.1103/PhysRevD.94.023516;%%
  %8 citations counted in INSPIRE as of 11 Feb 2017

%\cite{Belczynski:2016jno}
\bibitem{Belczynski:2016jno} 
  K.~Belczynski {\it et al.},
  %``The Effect of Pair-Instability Mass Loss on Black Hole Mergers,''
  Astron.\ Astrophys.\  {\bf 594}, A97 (2016),
  %doi:10.1051/0004-6361/201628980
  [arXiv:1607.03116 [astro-ph.HE]].
  %%CITATION = doi:10.1051/0004-6361/201628980;%%
  %5 citations counted in INSPIRE as of 11 Feb 2017


%  MORE DETECTIONS IN RUN O1

%\cite{Abbott:2016nmj}
\bibitem{Abbott:2016nmj} 
  B.~P.~Abbott {\it et al.} [LIGO Scientific and Virgo Collaborations],
  %``GW151226: Observation of Gravitational Waves from a 22-Solar-Mass Binary Black Hole Coalescence,''
  Phys.\ Rev.\ Lett.\  {\bf 116}, 241103 (2016),
  %doi:10.1103/PhysRevLett.116.241103
  [arXiv:1606.04855 [gr-qc]].
  %%CITATION = doi:10.1103/PhysRevLett.116.241103;%%
  %134 citations counted in INSPIRE as of 28 Sep 2016

%\cite{TheLIGOScientific:2016pea}
\bibitem{TheLIGOScientific:2016pea} 
  B.~P.~Abbott {\it et al.} [LIGO Scientific and Virgo Collaborations],
  %``Binary Black Hole Mergers in the first Advanced LIGO Observing Run,''
  Phys.\ Rev.\ X {\bf 6}, 041015 (2016),
  %doi:10.1103/PhysRevX.6.041015
  [arXiv:1606.04856 [gr-qc]].
  %%CITATION = doi:10.1103/PhysRevX.6.041015;%%
  %122 citations counted in INSPIRE as of 11 Feb 2017

%\cite{Abbott:2016bqf}
\bibitem{Abbott:2016bqf} 
  B.~P.~Abbott {\it et al.} [LIGO Scientific and Virgo Collaborations],
  %``The basic physics of the binary black hole merger GW150914,''
  Annalen Phys.\ {\bf 529}, 1600209 (2016),
  %doi:10.1002/andp.201600209
  [arXiv:1608.01940 [gr-qc]].
  %%CITATION = doi:10.1002/andp.201600209;%%
  %6 citations counted in INSPIRE as of 11 Feb 2017


%  STOCHASTIC BACKGROUND GW


%\cite{TheLIGOScientific:2016wyq}
\bibitem{TheLIGOScientific:2016wyq} 
  B.~P.~Abbott {\it et al.} [LIGO Scientific and Virgo Collaborations],
  %``GW150914: Implications for the stochastic gravitational wave background from binary black holes,''
  Phys.\ Rev.\ Lett.\  {\bf 116}, 131102 (2016),
  %doi:10.1103/PhysRevLett.116.131102
  [arXiv:1602.03847 [gr-qc]].
  %%CITATION = doi:10.1103/PhysRevLett.116.131102;%%
  %67 citations counted in INSPIRE as of 11 Feb 2017


%\cite{Mandic:2016lcn}
\bibitem{Mandic:2016lcn} 
  V.~Mandic, S.~Bird and I.~Cholis,
  %``Stochastic Gravitational-Wave Background due to Primordial Binary Black Hole Mergers,''
  Phys.\ Rev.\ Lett.\  {\bf 117}, 201102 (2016),
  %doi:10.1103/PhysRevLett.117.201102
  [arXiv:1608.06699 [astro-ph.CO]].
  %%CITATION = doi:10.1103/PhysRevLett.117.201102;%%
  %8 citations counted in INSPIRE as of 11 Feb 2017


%\cite{Clesse:2016ajp}
\bibitem{Clesse:2016ajp} 
  S.~Clesse and J.~Garc\'ia-Bellido,
  ``Detecting the gravitational wave background from primordial black hole dark matter,''
  arXiv:1610.08479 [astro-ph.CO].
  %%CITATION = ARXIV:1610.08479;%%
  %8 citations counted in INSPIRE as of 11 Feb 2017

 
% LISA 

%\cite{Bartolo:2016ami}
\bibitem{Bartolo:2016ami} 
  N.~Bartolo {\it et al.},
  %``Science with the space-based interferometer LISA. IV: Probing inflation with gravitational waves,''
  JCAP {\bf 1612}, 026 (2016),
  %doi:10.1088/1475-7516/2016/12/026
  [arXiv:1610.06481 [astro-ph.CO]].
  %%CITATION = doi:10.1088/1475-7516/2016/12/026;%%
  %7 citations counted in INSPIRE as of 11 Feb 2017


\bibitem{Bezrukov:2007ep} 
  F.~L.~Bezrukov and M.~Shaposhnikov,
  %``The Standard Model Higgs boson as the inflaton,''
  Phys.\ Lett.\ B {\bf 659}, 703 (2008),
 % doi:10.1016/j.physletb.2007.11.072
  [arXiv:0710.3755 [hep-th]].
  %%CITATION = doi:10.1016/j.physletb.2007.11.072;%%
  
  \bibitem{GarciaBellido:2008ab} 
  J.~Garc\'ia-Bellido, D.~G.~Figueroa and J.~Rubio,
  %``Preheating in the Standard Model with the Higgs-Inflaton coupled to gravity,''
  Phys.\ Rev.\ D {\bf 79}, 063531 (2009),
  %doi:10.1103/PhysRevD.79.063531
  [arXiv:0812.4624 [hep-ph]].
  %%CITATION = doi:10.1103/PhysRevD.79.063531;%%



%\cite{Allahverdi:2006iq}
\bibitem{Allahverdi:2006iq} 
  R.~Allahverdi, K.~Enqvist, J.~Garc\'ia-Bellido and A.~Mazumdar,
  %``Gauge invariant MSSM inflaton,''
  Phys.\ Rev.\ Lett.\  {\bf 97}, 191304 (2006),
  %doi:10.1103/PhysRevLett.97.191304
  [hep-ph/0605035].
  %%CITATION = doi:10.1103/PhysRevLett.97.191304;%%
  %258 citations counted in INSPIRE as of 12 Feb 2017

%\cite{Allahverdi:2006we}
\bibitem{Allahverdi:2006we} 
  R.~Allahverdi, K.~Enqvist, J.~Garc\'ia-Bellido, A.~Jokinen and A.~Mazumdar,
  %``MSSM flat direction inflation: Slow roll, stability, fine tunning and reheating,''
  JCAP {\bf 0706}, 019 (2007),
  %doi:10.1088/1475-7516/2007/06/019
  [hep-ph/0610134].
  %%CITATION = doi:10.1088/1475-7516/2007/06/019;%%
  %190 citations counted in INSPIRE as of 12 Feb 2017

%\cite{Linde:2007jn}
\bibitem{Linde:2007jn} 
  A.~D.~Linde and A.~Westphal,
  %``Accidental Inflation in String Theory,''
  JCAP {\bf 0803}, 005 (2008),
  %doi:10.1088/1475-7516/2008/03/005
  [arXiv:0712.1610 [hep-th]].
  %%CITATION = doi:10.1088/1475-7516/2008/03/005;%%
  %73 citations counted in INSPIRE as of 12 Feb 2017

%\cite{Hotchkiss:2011am}
\bibitem{Hotchkiss:2011am} 
  S.~Hotchkiss, A.~Mazumdar and S.~Nadathur,
  %``Inflection point inflation: WMAP constraints and a solution to the fine-tuning problem,''
  JCAP {\bf 1106}, 002 (2011),
  %doi:10.1088/1475-7516/2011/06/002
  [arXiv:1101.6046 [astro-ph.CO]].
  %%CITATION = doi:10.1088/1475-7516/2011/06/002;%%
  %32 citations counted in INSPIRE as of 12 Feb 2017


%\cite{Kannike:2017bxn}
\bibitem{Kannike:2017bxn} 
  K.~Kannike, L.~Marzola, M.~Raidal and H.~VeermŠe,
  %``Single Field Double Inflation and Primordial Black Holes,''
  arXiv:1705.06225 [astro-ph.CO].
  %%CITATION = ARXIV:1705.06225;%%
  %12 citations counted in INSPIRE as of 11 Sep 2017


\bibitem{Germani:2017bcs} 
  C.~Germani and T.~Prokopec,
  ``On primordial black holes from an inflection point,''
  arXiv:1706.04226 [astro-ph.CO].
  %%CITATION = ARXIV:1706.04226;%%

%\cite{Motohashi:2017kbs}
\bibitem{Motohashi:2017kbs} 
  H.~Motohashi and W.~Hu,
  %``Primordial Black Holes and Slow-Roll Violation,''
  Phys.\ Rev.\ D {\bf 96}, no. 6, 063503 (2017)
  doi:10.1103/PhysRevD.96.063503
  [arXiv:1706.06784 [astro-ph.CO]].
  %%CITATION = doi:10.1103/PhysRevD.96.063503;%%
  %4 citations counted in INSPIRE as of 11 Sep 2017

%\cite{Dimopoulos:2017ged}
\bibitem{Dimopoulos:2017ged} 
  K.~Dimopoulos,
  %``Slow-roll versus ultra slow-roll inflation,''
  arXiv:1707.05644 [hep-ph].
  %%CITATION = ARXIV:1707.05644;%%
  %3 citations counted in INSPIRE as of 11 Sep 2017


%\cite{Garcia-Bellido:2014gna}
\bibitem{Garcia-Bellido:2014gna} 
  J.~Garc\'ia-Bellido and D.~Roest,
  %``Large-$N$ running of the spectral index of inflation,''
  Phys.\ Rev.\ D {\bf 89}, 103527 (2014),
  %doi:10.1103/PhysRevD.89.103527
  [arXiv:1402.2059 [astro-ph.CO]].
  %%CITATION = doi:10.1103/PhysRevD.89.103527;%%
  %33 citations counted in INSPIRE as of 11 Feb 2017

  
%\cite{Bringmann:2011ut}
\bibitem{Bringmann:2011ut} 
  T.~Bringmann, P.~Scott and Y.~Akrami,
  %``Improved constraints on the primordial power spectrum at small scales from ultracompact minihalos,''
  Phys.\ Rev.\ D {\bf 85}, 125027 (2012),
  %doi:10.1103/PhysRevD.85.125027
  [arXiv:1110.2484 [astro-ph.CO]].
  %%CITATION = doi:10.1103/PhysRevD.85.125027;%%
  %69 citations counted in INSPIRE as of 10 Feb 2017

 % PLANCK

%\cite{Ade:2015lrj}
\bibitem{Ade:2015lrj} 
  P.~A.~R.~Ade {\it et al.} [Planck Collaboration],
  %``Planck 2015 results. XX. Constraints on inflation,''
  Astron.\ Astrophys.\  {\bf 594}, A20 (2016),
  %doi:10.1051/0004-6361/201525898
  [arXiv:1502.02114 [astro-ph.CO]].
  %%CITATION = doi:10.1051/0004-6361/201525898;%%
  %985 citations counted in INSPIRE as of 11 Feb 2017

%\cite{Ade:2015xua}
\bibitem{Ade:2015xua} 
  P.~A.~R.~Ade {\it et al.} [Planck Collaboration],
  %``Planck 2015 results. XIII. Cosmological parameters,''
  Astron.\ Astrophys.\  {\bf 594}, A13 (2016),
  %doi:10.1051/0004-6361/201525830
  [arXiv:1502.01589 [astro-ph.CO]].
  %%CITATION = doi:10.1051/0004-6361/201525830;%%
  %2831 citations counted in INSPIRE as of 11 Feb 2017

  
\bibitem{Harada:2013epa} 
  T.~Harada, C.~M.~Yoo and K.~Kohri,
  %``Threshold of primordial black hole formation,''
  Phys.\ Rev.\ D {\bf 88}, 084051 (2013), 
  Erratum: [Phys.\ Rev.\ D {\bf 89}, 029903 (2014)],
  %doi:10.1103/PhysRevD.88.084051, 10.1103/PhysRevD.89.029903
  [arXiv:1309.4201 [astro-ph.CO]].
  %%CITATION = doi:10.1103/PhysRevD.88.084051, 10.1103/PhysRevD.89.029903;%%

\bibitem{Green:2004wb} 
  A.~M.~Green, A.~R.~Liddle, K.~A.~Malik and M.~Sasaki,
  %``A New calculation of the mass fraction of primordial black holes,''
  Phys.\ Rev.\ D {\bf 70}, 041502 (2004),
  % doi:10.1103/PhysRevD.70.041502
  [astro-ph/0403181].
  %%CITATION = doi:10.1103/PhysRevD.70.041502;%%

\bibitem{Musco:2004ak} 
  I.~Musco, J.~C.~Miller and L.~Rezzolla,
  %``Computations of primordial black hole formation,''
  Class.\ Quant.\ Grav.\  {\bf 22}, 1405 (2005)
  %doi:10.1088/0264-9381/22/7/013
  [gr-qc/0412063].
  %%CITATION = doi:10.1088/0264-9381/22/7/013;%%

%\cite{Garcia-Bellido:2017aan}
\bibitem{Garcia-Bellido:2017aan} 
  J.~Garc\'ia-Bellido, M.~Peloso and C.~Unal,
  %``Gravitational Wave signatures of inflationary models from Primordial Black Hole Dark Matter,''
  arXiv:1707.02441 [astro-ph.CO].
  %%CITATION = ARXIV:1707.02441;%%
  %2 citations counted in INSPIRE as of 11 Sep 2017

%\cite{Carr:2017jsz}
\bibitem{Carr:2017jsz} 
  B.~Carr, M.~Raidal, T.~Tenkanen, V.~Vaskonen and H.~VeermŠe,
  %``Primordial black hole constraints for extended mass functions,''
  Phys.\ Rev.\ D {\bf 96}, no. 2, 023514 (2017)
  %doi:10.1103/PhysRevD.96.023514
  [arXiv:1705.05567 [astro-ph.CO]].
  %%CITATION = doi:10.1103/PhysRevD.96.023514;%%
  %19 citations counted in INSPIRE as of 12 Sep 2017

%\cite{Garcia-Bellido:2017fdg}
\bibitem{Garcia-Bellido:2017fdg} 
  J.~Garc\'ia-Bellido,
  %``Massive Primordial Black Holes as Dark Matter and their detection with Gravitational Waves,''
  J.\ Phys.\ Conf.\ Ser.\  {\bf 840}, no. 1, 012032 (2017)
  %doi:10.1088/1742-6596/840/1/012032
  [arXiv:1702.08275 [astro-ph.CO]].
  %%CITATION = doi:10.1088/1742-6596/840/1/012032;%%
  %16 citations counted in INSPIRE as of 12 Sep 2017
 
%\cite{Li:2016utv}
\bibitem{Li:2016utv} 
  T.~S.~Li {\it et al.} [DES Collaboration],
  %``Farthest Neighbor: The Distant Milky Way Satellite Eridanus II,''
  Astrophys.\ J.\  {\bf 838}, 8 (2017),
  %doi:10.3847/1538-4357/aa6113
  [arXiv:1611.05052 [astro-ph.GA]].
  %%CITATION = doi:10.3847/1538-4357/aa6113;%%
  %5 citations counted in INSPIRE as of 02 May 2017



\end{thebibliography}
\end{document}